\documentclass[preprint,showpacs,preprintnumbers,amsmath,amssymb]{revtex4}
\usepackage{graphicx}
\usepackage{dcolumn}
\usepackage{bm}
\usepackage{epsfig}
\usepackage{subfigure}

\newcommand{\bi}{\begin{itemize}}
\newcommand{\ei}{\end{itemize}}
\newcommand{\be}{\begin{eqnarray}}
\newcommand{\ee}{\end{eqnarray}}
\newcommand{\bbmatrix}{\left( \begin{array}}
\newcommand{\eematrix}{\end{array} \right)}

\begin{document}

\title{Theory of Manganite Superlattices}
\author{Chungwei Lin and Andrew.~J.~Millis}
\affiliation{ Department of Physics, Columbia University \\
538W 120th St NY, NY 10027}

\begin{abstract}
A theoretical model is proposed for the (0,0,1) superlattice manganite system (LaMnO$_3$)$_n$(SrMnO$_3$)$_m$.
The model includes the electron-electron, electron-phonon, and cooperative Jahn-Teller interactions.
It is solved using a version of single-site the dynamical mean field approximation generalized to
incorporate the cooperative Jahn-Teller effect. The phase diagram and conductivities are calculated.
The behavior of the superlattice is found to a good approximation to be
an average over the density-dependent properties of individual layers, with the density 
of each layer fixed by electrostatics.
\end{abstract}
\maketitle


\section{Introduction}
An exciting recent development in materials science is the ability to fabricate atomically precise multilayer structures
involving transition metal oxides \cite{Ohtomo_02,Chakhalian_06,Brinkman_07, Adamo_08, May_07,Bhattacharya_07}. 
Multilayer structure based on simple semiconductors
such as Si and GaAs/AlGaAs give rise to wide range of striking physical phenomena such as the integer and fractional
quantum Hall effects \cite{QF1,QF2}  and are important for many classes of devices. It is therefore natural to expect that
transition metal oxides, which display a far richer variety of phenomena than do semiconductors \cite{Imada_98}, will in heterostructure form
yield an even more diverse set of new phenomena. 

The colossal magnetoresistance (CMR) rare earth manganites La$_{1-x}$Sr$_x$MnO$_3$ have been studied intensively
\cite{Goodenough_55, CMR, Millis_96, AhnMillis_00, Michaelis_03, cLin_08} and provide an important model system. These 
compounds display a wide range of phases with characteristic orbital, magnetic and transport signatures
\cite{Goodenough_55, Wollan_55, Schiffer_95, CMR}, providing opportunities both 
for creating many effects in the superlattice context and for detecting phases that may be created. Manganite
superlattices have now been fabricated and are being studied experimentally \cite{Adamo_08, May_07,Bhattacharya_07} but
have not yet received much theoretical attention. In 
this paper we present a theoretical analysis of these interesting systems. Our findings may be summarized by the
following 4 rules: (1) the layer {\em charge distribution} mainly depends on the electrostatic interaction and has very
weak dependence on the temperatures and on whether or not the system has orbital or magnetic order; 
(2) once the layer charge distribution is given, the {\em orbital order}
at each layer is essentially determined by the bulk behavior at the same density: the propagation length of the orbital order
along the superlattice is less than a lattice constant; (3) the inter-layer {\em magnetic coupling} is ferromagnetic (FM)
except between layers with densities close to 1, where it is anti-ferromagnetic (AFM); (4) the {\em in-plane conductivity} 
is essentially the average of {\em bulk} conductivities for the given layer charge distribution. 
These results indicate that the main effect of superlattice structure is simply to produce a layer-dependent charge distribution.
Once the layer densities are known, each layer behaves basically according to the bulk phase diagram for that density
with only weak proximity effects.

The rest of this paper is organized as follows. First we define the superlattice and discuss the interactions 
considered in the bulk model, then we present and discuss results for a particular superlattice consisting of
four layers of LaMnO$_3$ and one of SrMnO$_3$ in detail. From these and related calculations we infer the rules outlined above.
Finally we use the rules to discuss recent experiments \cite{Adamo_08, May_07} arguing that the data still
may not be displaying intrinsic behavior. We propose experiments on a variant of the superlattice which may be
more definitive.

\section{Definitions, Hamiltonian, Methods, and Phases}

\subsection{Definition of Superlattice}
\begin{figure}[htbp]
   \centering
   \epsfig{file= 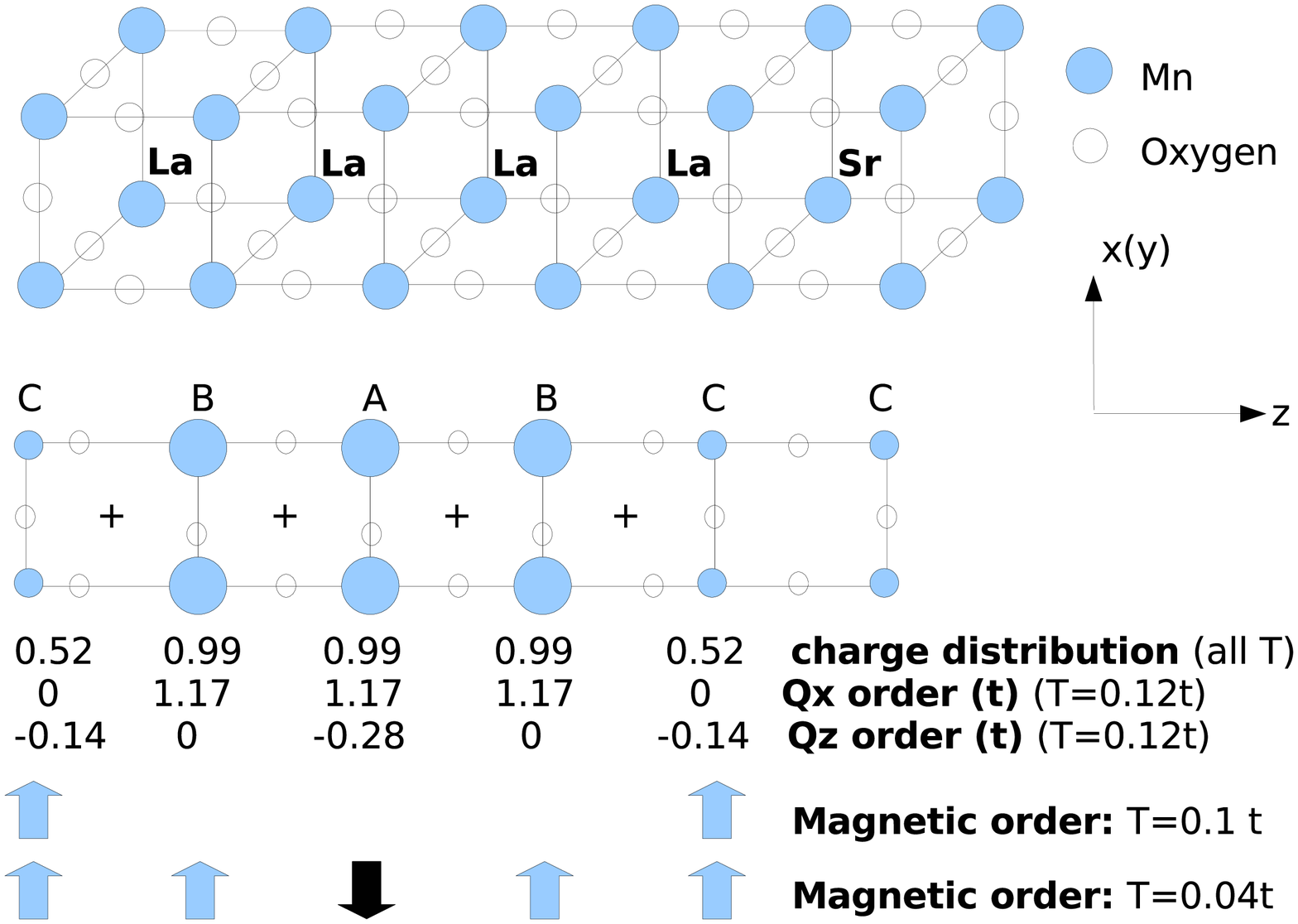, width=0.5 \textwidth}
   \caption{Upper panel: Schematic representation of a (LaMnO$_3$)$_4$ (SrMnO$_3$)$_1$ superlattice
	with Mn manganese and O oxygen ions denoted by 	shaded and open circles respectively. 
	Lanthanum and Strontium ions labeled by La/Sr reside in the center of cubes defined by Mn.
	Lower panel: Upper part -- Projection of (4,1) superlattice onto $x-z$ plane, with the three symmetry
	inequivalent Mn sites labeled A,B,C.  Middle part -- a numerical representation of charge density and amplitude of
	orbital order computed for screening parameter $\alpha=0.3$ at temperature indicated.  Lower part -- pictorial representation
	of magnetic order (order parameter direction indicated by arrows)  calculated using screening
	parameter $\alpha=0.3$ for  two temperatures as indicated. 
	}
   \label{fig:superlattice}
\end{figure}

The high temperature structure of LaMnO$_3$ is a slightly distorted version of the ABO$_3$ perovskite. It
may be thought of as a simple cubic lattice with Mn on the cube vertices and La in the body centers.
The superlattices of interest here are composed of the same
cubic lattice of Mn, with a superstructure defined by a periodic  
replacement of La by Sr. To specify the superlattice we must specify the direction along which  
the $La/Sr$ sites alternate and the way the $La$ and $Sr$ ions are arranged. We
will consider a (0,0,1) (LaMnO$_3$)$_m$(SrMnO$_3$)$_n$ superlattice, abbreviated as L$m$S$n$ or ($m,n$), 
defined by $m$ La planes perpendicular to the (0,0,1) cube axis, followed by $n$ Sr planes. The entire structure has
a periodicity $(m+n)$ in the $z$ (coordinate) direction while maintaining the original lattice  translational symmetry
in $x$ and $y$. The upper panel of Fig(\ref{fig:superlattice}) shows a L4S1, (4,1) superlattice.
We shall refer to each $x-y$ plane as a layer.

\subsection{Hamiltonian and Parameters}
The degrees of freedom needed to describe the physics of manganites are \cite{CMR} the two $e_g$-like orbitals
which make up Mn conduction band, an electrically inert classical core spin 
representing electrons occupying $t_{2g}$ orbitals at Mn sites, and the three even-parity MnO$_6$ distortion modes 
$Q_0$, $Q_x$, $Q_z$. The Hamiltonian for the manganite superlattice is the Hamiltonian for bulk manganites,
supplemented by Coulombic terms representing the potential arising from the pattern of the La and Sr ions,
thus $H=H_{Bulk} + H_{Coul}$. The bulk Hamiltonian is
$H_{Bulk} = H_{band}+H_{EE} + H_{Hund}+H_{el-lat} + H_{lattice}$ where
\begin{eqnarray}
H_{band} 
= \sum_{\vec{k}, ab, \sigma} \epsilon_{\vec{k}, ab, \sigma} c^{\dagger}_{\vec{k}, a,\sigma}  c_{\vec{k},b,\sigma}
\label{eqn:H_band} 
\end{eqnarray}
with $\epsilon_{\vec{k}, ab, \sigma} = -t (\varepsilon_0 \hat{e} + \varepsilon_z \hat{\tau}_z+ \varepsilon_x \hat{\tau}_x)_{ab}$
where $e$ is the unit matrix, $\hat{\tau}$ Pauli matrices and $\varepsilon_0 = \cos k_x + \cos k_y + \cos k_z $, 
$\varepsilon_z = \cos k_z -\frac{1}{2} (\cos k_x + \cos k_y )$, and
$\varepsilon_x = \frac{\sqrt{3}}{2} (\cos k_x - \cos k_y)$.
$a,b$ label orbitals, $i,j$ sites, and $\sigma$ spins.
From our previous band structure calculation \cite{Ederer_07}, the hopping parameter $t$ is $0.65$eV which defines
the energy unit. 
\begin{eqnarray}
H_{EE} &=& (U- J) n_1 n_2 + U \sum_{
i=1,2}n_{i,\uparrow} n_{i, \downarrow } + J( \, c^{\dagger}_{1,
\uparrow} c^{\dagger}_{1, \downarrow} c_{2, \downarrow} c_{2,
\uparrow} +h.c.)- 2 J \vec{S}_1 \cdot \vec{S}_2 
\label{eqn:H_EE}
\end{eqnarray}
with $\vec{S}_{1 (2)} = \vec{\sigma}_{\alpha \beta} c^{\dagger}_{1(2),\alpha} c_{1(2),\beta}$.
The values $U\sim2.5$eV and $J\sim 0.5$eV are found to be appropriate for the manganite \cite{cLin_08-2}.
\begin{eqnarray}
H_{Hund} &=& -J_H \sum_{i}\vec{S}_i \cdot c^{\dagger}_{i,\alpha} \vec{\sigma}_{\alpha \beta}
 c_{i,\beta} 
\label{eqn:H_Hunds} 
\end{eqnarray}
with $J_H\sim1.5$eV and $|\vec{S}| = 1$. An additional antiferromagnetic core-spin core-spin interaction
exists in the material. This interaction is crucial for $x>0.5$ but affects only minor details
of the $x<0.5$ regime of interest here, i.e. it changes the phase boundaries but does not
lose any exhibited phases. Including this interaction involves too great a computational
expense, so we do not include it here. 

For the electron-lattice interactions \cite{Millis_96} we only include the Jahn-Teller coupling
\be
H_{JT} = -\sum_{i,a,b}  ( Q_{i,x} \tau^x_{ab} + Q_{i,z} \tau^z_{ab})
 c^{\dagger}_{i,a} c_{i,b} 
= -\sum_{i,a,b} \vec{Q}_i \cdot \vec{\tau}_{ab} c^{\dagger}_{i,a} c_{i,b}
\label{eqn:H_JT0} 
\ee
with $\vec{Q}_i=(Q_{i,x},Q_{i,z}) = Q_i(\sin 2\theta_i, \cos 2\theta_i)$ ($0<\theta<\pi$). 
We neglect the breathing mode coupling \cite{Millis_96} in the current calculation.

For the lattice Hamiltonian we include an elastic energy term from adjacent Mn-O and Mn-Mn force constants
\cite{Millis_96-2, cLin_08} and the cubic term noted by Kanamori \cite{Kanamori_61} 
\be
H_{lattice}= \sum_{ij,ab}\frac{1}{2K^{ab}_{ij} } Q_{i,a} Q_{j,b} -A \sum_i (3 Q^3_{i,z}- Q^2_{i,x}Q_{i,z})
\ee
with $i,j$ labeling site while $a,b$ distortion modes.

As far as $e_g$ electrons are concerned, La and Sr ions act as +1 and neutral point charges respectively \cite{Okamoto_04,cLin_06}. 
In the superlattice, the distribution of those cations is patterned, and the Hamiltonian from electrostatics is
\begin{eqnarray}
H_{Coul}= \sum_{ij} \left[ \frac{1}{2\bar{\epsilon}} \frac{e^2 n_i n_j}{|\vec{r}_i-\vec{r}_j|}
+\frac{1}{2\bar{\epsilon}} \frac{e^2}{|\vec{R}^{La}_i-\vec{R}^{La}_j|}
-\frac{e^2 n_i}{\bar{\epsilon}|\vec{r}_i-\vec{R}^{La}_j|} \right]
\end{eqnarray}
with $n_i=\sum_{\sigma, a}c_{i,a,\sigma}^+ c_{i,a,\sigma}$ the occupation number
at Mn site $i$. $\vec{r}_i$ and $\vec{R}^{La}_i$ are positions of $Mn$ and $La$ in $i$th unit cell, 
and $\bar{\epsilon}$ is the dielectric constant of the material.
To describe the magnitude of this interaction, we define the dimensionless screening parameter $\alpha \equiv e^2/a \bar{\epsilon} t$ 
which controls the charge density distribution. 
The value of $\alpha$ has not been determined, but its order of magnitude may be
estimated from the hopping parameter $t\sim 0.65$eV, lattice constant $a=4$\AA $ $
and typical value of dielectric constant $\bar{\epsilon}\sim10$ \cite{Okamoto_06} to be $\alpha = 0.2$. 

\subsection{Method and Results from Bulk Model}

We now briefly discuss the method of solving the model. To solve the model we use the single-site
dynamical mean field theory (DMFT) \cite{DMFT_96} with semiclassical impurity solver \cite{Okamoto_05}. 
The intersite lattice coupling (cooperative Jahn-Teller effect \cite{Halperin_71, Millis_96-2}) is taken into account
within the single-site DMFT by the mean field approximation, i.e. the local impurity problem contains an effective
interaction from distortions at neighboring sites \cite{cLin_08}. 
For the term introduced by the superlattice $H_{Coul}$, we adopt the Hartree approximation whose overall effect is to
generate a self-consistent potential at each Mn site \cite{Okamoto_04, cLin_06}.
The optical conductivities are computed in the same manner as described in Ref\cite{Michaelis_03}, and the DC
values are obtained by taking the $\omega \rightarrow 0$ limit.
Sepcifics of the DMFT computation are presented in Ref\cite{cLin_08}.

\begin{figure}[htbp]
\centering
   \epsfig{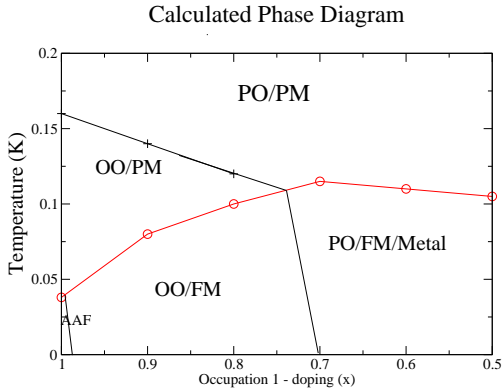} 
   \caption{The calculated bulk phase diagrams. To make comparison between our calculations
	and experiments, the calculation done at $T=0.1t$ correspond to roughly 250$K$ in the experiments \cite{cLin_08}.
	}
   \label{fig:BulkPhaseDiagram}
\end{figure}

The first step to study the superlattice is to understand the parental bulk system. 
Fig(\ref{fig:BulkPhaseDiagram}) shows the theoretically calculated bulk phase diagram. 
Fig(\ref{fig:BulkPhaseDiagram}) is different from what we presented in Ref\cite{cLin_08} in phase boundaries 
because here we neglect the AF magnetic coupling between Mn core spins. 
This simplification does not lose any phases for the doping range $x < 0.5$ but does lose phases for $x > 0.5$, 
therefore we only present bulk results for $x<0.5$ and only consider $(n(\geq 2),1)$ superlattices where all layers have  
$e_g$ electron occupations larger than 0.5.
Our bulk calculation overestimates the transition temperatures but gives a reasonable description of the ordering 
of phases. For this reason we focus on the exhibited phases in superlattice but not the transition temperatures on the numerical values.

\subsection{Planar Long-ranged Orders}
\begin{figure}[htbp]
   \centering
   \epsfig{file=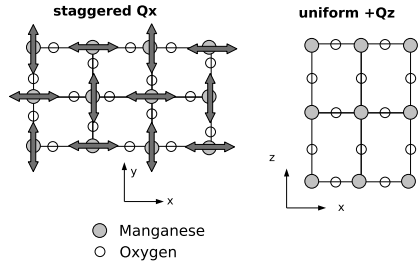,width=0.4\textwidth}
   \caption{ The orbital orders. (a) The staggered $Q_x$ order and (b) the uniform $Q_z$ order.
	For the staggered $Q_x$ order, all octahedra in the plane undergo $Q_x$ distortions with the long-axis
	(double arrows) of each distorted octahedron differs by 90 degree from its adjacent sites. 
	For the +(-)$Q_z$ order, all oxygens are in the middle of neighboring Mn, but
	the Mn-O distance in $z$ direction is longer(shorter) than that in $x$ and $y$.
	}
   \label{fig:orbitalorder}
\end{figure}

Similar to earlier findings \cite{Okamoto_04, cLin_06}, the superlattice results are most naturally
presented by first describing the charge density and exhibited order(s) at each $x-y$ plane, and the superlattice phase
are therefore the planer phases plus the relations between them. 
To facilitate later discussion, we summarize here the three planar long-ranged orders found
in our calculation: a ($\pi,\pi$) staggered $Q_x$, a uniform $Q_z$, and a ferromagnetic (FM) order.
The first two are orbital orders; the ordering pattern are shown in Fig(\ref{fig:orbitalorder}).
For the staggered $Q_x$ order, all octahedra in the plane undergo $Q_x$ distortions with the long-axis
of each distorted octahedron differs by 90 degree from its adjacent ones, as shown in Fig(\ref{fig:orbitalorder})(a). 
For the +(-)$Q_z$ order, all oxygens are in the middle of neighboring Mn, but
the Mn-O distance in $z$ direction is longer(shorter) than that in $x$ and $y$, as Fig(\ref{fig:orbitalorder})(b). 

An equivalent way to describe the local orbital order is to use the ground state 
$| \theta \rangle = \cos \theta |3z^2-r^2 \rangle + \sin \theta |x^2-y^2 \rangle$ \cite{cLin_08} of the
Jahn-Teller coupling (Eq(\ref{eqn:H_JT0})) with $\vec{Q}_i$ given by the distortions.
In this language the ($\pi,\pi$) $Q_x$ order corresponds to an alternating $|\theta=\pi/4 \rangle$,  
$|\theta=-\pi/4 \rangle$ orbital configuration and the uniform $+(-)Q_z$ a uniform  $|\theta=0\rangle$ ($|\theta=\pi/2 \rangle$).


\section{Results for $(4,1)$ Superlattice}
In this section we present our calculated results for the $(4,1)$ superlattice, a representative case that captures most of the 
relevant phenomena. We consider two different values of $\alpha$; $\alpha=0.09$ corresponding to relatively
delocalized electrons and $\alpha=0.3$ corresponding to tightly confined electrons.

\begin{figure}[ttt]
\begin{center}
   \epsfig{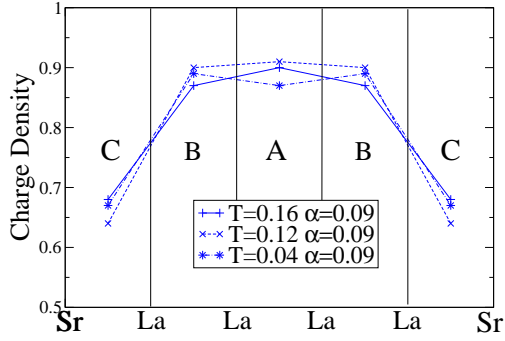}
   \caption{L4S1 charge distribution for $\alpha=0.09$ at $0.16t$ (above any ordering temperatures), 
	$0.12t$ (orbital ordering but not magnetic ordering in superlattices), and $0.04t$
	(both orbital and magnetic order presented).
	}
   \label{fig:L4S1profile}
\end{center}
\end{figure}

For L4S1 each supercell has 3 symmetry-inequivalent layers, labeled as A B C where A is sandwiched by La, B by La and Sr, and
C by Sr, as illustrated in Fig(\ref{fig:superlattice})(b).
Fig(\ref{fig:L4S1profile}) presents the L4S1 layer charge distribution 
for $\alpha = 0.09$ (corresponding to a relatively delocalized electrons) at temperatures 
$0.16t$ (above any ordered temperatures), $0.12t$ (orbital ordering but not magnetic ordering presented), and $0.04t$
(both orbital and magnetic order). The variation in charge distribution for different temperatures is smaller than 5$\%$.

%

{\em The staggered $Q_x$ order}:
Table I lists the layer charge density, the computed SL $Q_x$ order, and the bulk $Q_x$ order at the given 
layer charge density of the (4,1) SL for $\alpha=0.3$ and $0.09$ at $T=0.12t$. 
We see that the SL calculation is very close to the bulk ($<1\%$) which means that the $Q_x$ order is mainly determined by 
its layer charge density and the proximity effect is very weak. 

\begin{center}
\begin{tabular}{|l||c|c|c||c|c|c|} \hline
 & \multicolumn{3}{c||}{$\alpha$=0.3} & \multicolumn{3}{c|}{$\alpha$=0.09} \\ \hline
site label              & A     &  B  &  C  &  A  &  B  &  C  \\ \hline
Charge Density          & 0.99  & 0.99 & 0.52 & 0.92 & 0.90 & 0.64 \\ \hline
$| \langle Q^s_x \rangle |$ SL $[t]$    & 1.17 & 1.17 & 0 & 1.01 & 1.0 & 0 \\ \hline
$| \langle Q^s_x \rangle |$ Bulk  $[t]$  & 1.18 & 1.18 & 0 & 1.02 & 1.0 & 0 \\ \hline
$| \langle Q^u_z \rangle |$ SL $[t]$    & -0.28 &  0  &  -0.14 & -0.18 & -0.03 & -0.08 \\ \hline
$| \langle Q^u_z \rangle |$ Bulk $[t]$  & -0.32 & -0.32 & 0 & -0.22 & -0.2 & 0 \\ \hline
\end{tabular} \\
Table I: The amplitudes of the staggered $Q_x$ ($| \langle Q^s_x \rangle |$), 
uniform $Q_z$ ($| \langle Q^u_z \rangle |$)\\ orders for the L4S1 superlattice  and for the corresponding
bulk values \\ at same densities for $T=0.12t$ \\
\end{center}
The very weak orbital order proximity effect for the staggered $Q_x$ order has a straightforward physical origin.
Rotation symmetry about the bond in the $z$ direction means that neither nearest neighbor (NN) hopping nor nearest
neighbor atom-atom forces can distinguish $+Q_x$ order from $-Q_x$ order. One consequence is that within 
this approximation $(\pi,\pi,\pi)$ $Q_x$ and $(\pi,\pi,0)$ $Q_x$ order are degenerate implying that the in-plane
$(\pi,\pi)$ $Q_x$ order does not have any dispersion in $z$ direction and thus no inter-layer coupling
(no proximity effect) for this order. Second neighbor hopping \cite{cLin_08}
or certain classes of shear elastic forces \cite{AhnMillis_01} do distinguish the two but the differences are small (of the order of meV).
We will give a more detailed discussion in the discussion section.


{\em The uniform $Q_z$ order}:
The 3rd and 4th rows in Table I list the  computed SL $Q_z$ order of the (4,1) SL for $\alpha=0.3$ and 0.09 at $T=0.12t$
and the corresponding bulk results at the same density.
These results can be understood as follows. There are two apparent sources affecting the uniform $Q_z$ order -- 
the anharmonic term in $H_{lattice}$ and the SL effect. The net SL effect is to produce an electrostatic like force, i.e.
oxygens are pushed toward to Mn layer with less $e_g$ density. One bears in mind that the origin of this force
is because the octahedral distortion changes the Mn-O hybridization thus affects the 
population distribution of $e_g$ orbitals, and has little to do with the actual static Coulomb energy \cite{cLin_08}.
For the parameters considered above, 
Layer A is sandwiched by Layer B having similar charge densities (1 in this case), therefore the SL effect is weak for A and
the main contribution is from the anharmonic effect. Indeed the $Q_z$ order of Layer A is -0.28$t$, very close to the bulk result 
-0.32$t$. For layer B, besides the bulk effect which causes a -$Q_z$ order, the charge inhomogeneity results 
in a +$Q_z$ order since oxygens are pushed {\em away} from layer B and the in this case these two effects nearly cancel each other. 
For layer C, only the SL effect contributes and the result is a 0.14$t$ $-Q_z$ order. We see that at interfaces the bulk and SL effects
are of comparable strengths therefore the uniform $Q_z$ order has significant proximity effect.

{\em The magnetic order}:
Now we discuss the magnetic order for $\alpha=0.3$ and 0.09.
For $\alpha=0.3$, only layer C has FM order at $T=0.1t$. When lowering the temperature to $0.04t$, all layers are 2D FM. The
interlayer coupling between A and B (with layer densities roughly one) is AF, and all others are FM as shown in Fig(\ref{fig:superlattice})(b).
For $\alpha=0.09$, one obtains a smoother charge distribution and
at low temperature all inter-layer magnetic couplings are FM. 
Those results suggest that the interlayer magnetic coupling is generally ferromagnetic except that between layers with densities close to one.
The FM coupling between layers can be understood from double exchange mechanism where FM arrangement maximizes
the kinetic energy and is energy favored when the conduction band is not full. 
The AF coupling between $N=1$ layers has the same origin as the bulk \cite{cLin_08} where the system gains gap energy
by arranging core spins antiferromagnetically. 
One interesting consequence of these results is that if the charge density is sufficiently sharp enough
to introduce AF inter-layer coupling ($\alpha=0.3$ for example), the superstructures with odd number 
of La have an odd number of AF bonds so the magnetic lattice has twice the period of the structured one. 
For example for the L3S1 superlattice, we expect the period including the magnetic order is 8 and adding up magnetic orders from all layer 
leads to zero total magnetization.


\section{Spectral Functions and Conductivities for Different Superlattices}
In this section we shall first establish the rule implied from our superlattice calculation for the in-plane conductivity and
also discuss the out-of-plane one. Based on the established rule we discuss the temperature and
superlattice dependence on the DC measurements.

\subsection{The rule for the conductivity}
\begin{figure}[htbp]
\centering
   \subfigure[]{\epsfig{file = 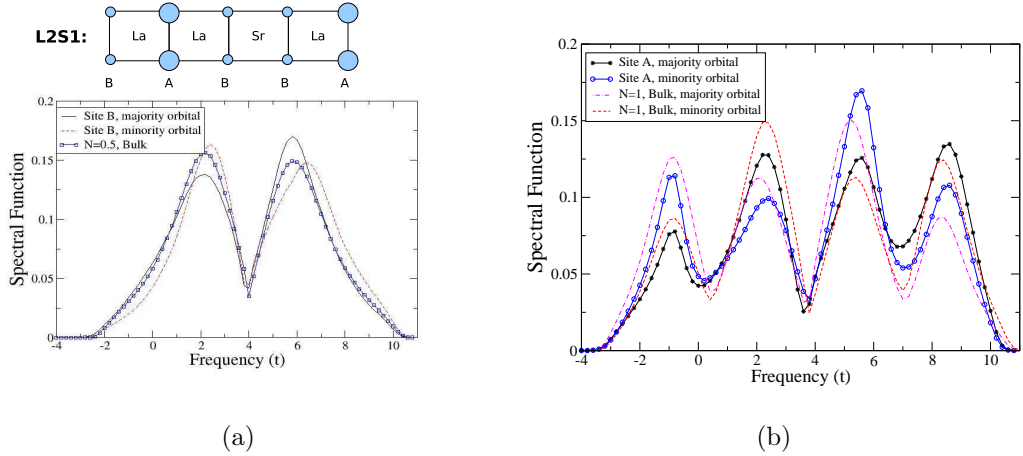, width=0.45 \textwidth}} 
   \subfigure[]{\epsfig{file = DOS_L2S1_SiteA_vs_PiPi0_t1_eps0.20_Cubic_0.020_UQ1.4_JH2.0_Us1.2_Occu1.0_Temp0.120_BulkShifted.eps, width=0.4\textwidth}} 
   \caption{Spectral functions for the (2,1) superlattice compared to bulk spectra at density equal to layer density for
	$\alpha=0.3$ and at $T=0.12t$. (a) The superlattice spectral functions for layer B and
	bulk at $N=0.5$. (b) SL spectral functions for layer A and bulk at $N=1$. 
	To facilitate the comparison, the bulk results are shifted 0.4$t$ in frequency.  
	The peak-peak distances are essentially the same for bulk and SL calculations. 	
	}
   \label{fig:Bulk_SL_DOS}
\end{figure}

We found that once the layer charge densities $n_i$ are specified, the excitation spectrum at each layer behaves very close to
the bulk at the same density. We use the simplest superlattice L2S1 to demonstrate this point. For (2,1) SL there are
only two symmetry-inequivalent sites, one between La labeled as A while the other between La and Sr B, as shown in
Fig(\ref{fig:Bulk_SL_DOS}) (a). For $\alpha=0.3$ and $T=0.12t$ we found $n_A \sim 1$ and $n_B\sim 0.5$. 
Fig(\ref{fig:Bulk_SL_DOS}) (a)/(b) shows both the calculated SL spectral functions for layer B(A) and the bulk for $N=0.5$($N=1$)
at the same temperature. 

\begin{figure}[htbp]
\centering
   \epsfig{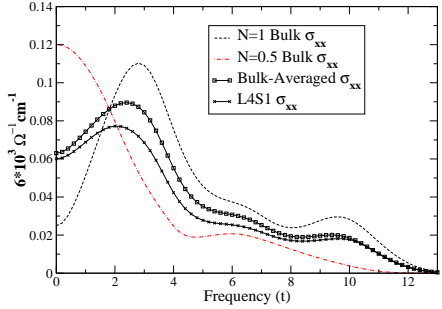} 
   \caption{Calculated SL in-plane optical conductivity and the corresponding bulk-averaged one for the L4S1 at $\alpha=0.3$, $T=0.12t$. 
	As a reference we also plot the bulk $\sigma_{xx}(\omega)$ for $N=1$ and $N=0.5$ (curves without ligands)
	at the same temperature which are used for computing the bulk-averaged conductivities. 
	}
   \label{fig:bulkaverage}
\end{figure}

Since the in-plane conductivity $\sigma_{xx}^{SL}$ is dominated by in-plane hopping processes 
(moving electrons from one site to the other within the same layer) $and$
the spectral function for layer with density $n_i$ is very close to the bulk at the same density,
$\sigma_{xx}^{SL}$ should be very close to the bulk-averaged value $\bar{\sigma}_{xx}^{bulk}$ which is define as
\be
\bar{\sigma}_{xx}^{bulk} \equiv \sum_{i=1}^N \sigma_{xx}^{bulk} (n_i) /N
\label{eqn:bulk_average}
\ee
To see how accurate this rule is, we list in
Table III the calculated SL DC conductivity $\sigma_{xx}^{SL}$ and the bulk averaged $\bar{\sigma}_{xx}^{bulk}$ 
according to the obtained charge distribution at $T=0.12t$  where 
the unit for $\sigma_{xx}$ is $e^2 / (\hbar a) \sim 6 \times 10^3 (\Omega cm)^{-1}$ with $a$ the lattice constant ($\sim 4$\AA).
\begin{center}
\begin{tabular}{|l|l|l||l|l||l|l|} \hline
$\alpha$              & $0.3$ & $0.09$ & $0.3$ & $0.09$ & $0.3$ & $0.09$ \\ \hline
$\bar{\sigma}_{xx}^{bulk}$ & 0.083 & 0.115 & 0.064 & 0.093 & 0.58 & 0.082 \\ \hline
$\sigma_{xx}^{SL}$         & 0.077 & 0.105 & 0.067 & 0.093 & 0.6 & 0.081 \\ \hline
SL & L2S1 & L2S1    & L3S1 & L3S1 & L4S1 & L4S1 \\ \hline
\end{tabular} \\
Table II $\sigma_{xx}^{SL}$ and $\bar{\sigma}_{xx}^{bulk}$ for different superlattices \\
\end{center}
As a concrete example we compute bulk-averaged $\bar{\sigma}_{xx}^{bulk}$ for L4S1, $\alpha=0.09$, $T=0.12t$ in details. From 
Fig(\ref{fig:L4S1profile}) we see that there are two layers with density 0.9 (layer B), two layers 0.64 (layer C), and one
layer 0.91 (layer A) for those parameters. Now we consult the bulk results for DC conductivity at $T=0.12t$
and find that $\sigma_{xx}^{bulk}$ at $N=$0.9, 0.64, 0.91 are 0.055, 0.12, 0.05 respectively.
The bulk averaged $\bar{\sigma}_{xx}^{bulk}$ is thus $(2 \sigma_{xx}^{bulk}(n=0.9)+ 2\sigma_{xx}^{bulk}(n=0.64)+\sigma_{xx}^{bulk}(n=0.91))/5$ roughly
0.081. As listed in Table II, the difference between $\bar{\sigma}^{bulk}$ and $\sigma_{xx}^{SL}$ 
is roughly $10\%$ for L2S1, and is smaller ($\sim 2\%$) for L3S1 and L4S1.
Fig(\ref{fig:bulkaverage}) compares the in-plane optical conductivities with the bulk-averaged for the L4S1 at $\alpha=0.3$, $T=0.12$.
We see they match quite well $(\lesssim 10\%)$ (both peak positions and peak amplitudes) for all frequencies.

Unlike the in-plane case, the out-of-plane conductivity $\sigma_{zz}^{SL}$ involves inter-layer hopping processes
(removing electrons from layer $i$ and adding them to layer $i\pm \hat{z}$). Since different layers experience different
static Coulomb potential, there is no simple relation between $\sigma_{zz}^{SL}$ and $\{n_i\}$. 
Our previous study \cite{cLin_06} of a simpler model system, namely the one-orbital double-exchange superlattice, shows that
there are peaks in $\sigma_{zz}^{SL}(\omega)$ directly corresponding to the potential difference between layers
from which the $\alpha$ is determined straightforwardly. However for the current problem the potential difference
is mainly to produce layer-dependent local spectra and are not directly related to peaks in $\sigma_{zz}^{SL}$.
Nonetheless $\sigma_{zz}^{SL}$ is more sensitive to the screening parameter than $\sigma_{xx}^{SL}$ and
can be used to constrain the values of $\alpha$. We will return to this point in the discussion section.

\subsection{Temperature dependence of DC resistivity}
\begin{figure}[htbp]
\centering
   \epsfig{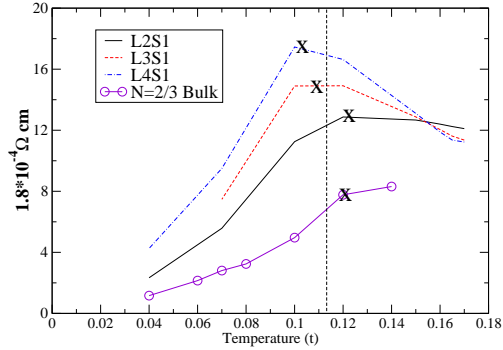} 
   \caption{The DC resistivity as a function of temperature for $\alpha=0.3$. $X$ indicates the
	calculated interface Curie temperature for each superlattice.
	The vertical dashed line marks a rough temperature where $d\rho/dT$ changes sign.
	As a reference we also present the bulk result for $N=2/3$ ($x=1/3$). 
	}
   \label{fig:DCResistivity(T)}
\end{figure}

Now we discuss the temperature dependence on the DC resistivity $\rho$. Fig(\ref{fig:DCResistivity(T)})
shows $\rho(T)$ for $\alpha=0.3$. We observe that around $T=0.11t$ $d\rho/dT$ changes sign implying the superlattice
goes from high T insulating phase to low T metallic. This $downturn$ in $\rho(T)$ coincides with
the interface Curie temperatures which are also marked Fig(\ref{fig:DCResistivity(T)}). 
Across this temperature the interfaces go from PM/Bad Metal to FM/Metal accounting
for the superlattice metallic behavior. From double-exchange mechanism \cite{Michaelis_03, cLin_08}, the PM/FM transition is 
always accompanied with a insulator/metal (bad metal/metal at least) transition, the sign change in $d\rho/dT$ 
as a function of temperature is quite general for all values of $\alpha$, but is more pronounced for large $\alpha$.

\subsection{Superlattice dependence of DC conductivity}
Since for small $\alpha$, the superlattice behaves just like the bulk material, we only focus on the $\alpha$
which leads to the sharp charge distribution, i.e. $\alpha=0.3$. Our result implies that the SL conductivity
is proportional to the interface density at low temperature (the interface density is defined as the ratio between
the number of Mn sandwiched by La $and$ Sr and that of total Mn layers, for L2S1 the interface density is 2/3,
for L4S1 5/2). This statement is a direct consequence of the rule $\sigma_{xx}^{SL} \sim \bar{\sigma}_{xx}^{bulk}$:
for sharp charge distribution, $N\sim 1$ layers are insulating at low T and only interface layers ($N\sim 0.5$) are conducting.
To demonstrate this statement from the SL calculation, we define $r(T) = \sigma_{xx}^{L4S1}(T)/\sigma_{xx}^{L2S1}(T)$ and compare it
with the ratio of interface densities between L4S1 and L2S1 which is 0.6. The following table
presents $r(T)$ for several temperatures:
\begin{center}
\begin{tabular}{|l||l|l|l|l|} \hline
$T/t$ & $0.15$ & $0.1$ & $0.07$ & $0.04$  \\ \hline
$r(T)^{\alpha=0.3}$   & 0.96 & 0.65 & 0.55 & 0.57 \\ \hline
\end{tabular} 
\end{center}
We see that $r(T)$ is indeed very close to 0.6 below the interface Curie temperature ($\sim 0.11t$). 
Finally we emphasize that from our calculation, the statement ``SL conductivities are only from interfaces''
is true only below temperature $and$ with sharp charge distribution.

\section{Discussion}
In this section we first summarize our results by providing rules deduced from our calculations, then discuss
how to determine the layer charge distribution and thus the screening parameter experimentally. Finally
we give a more detailed analysis on the proximity effect of the orbital orders.
\subsection{Rules implied by the calculation}
The exhibited phases at each layers are the combined effects of bulk Hamiltonian and the charge inhomogeneity
induced by the SL. Here we summarize our calculations by stating the following rules which governs
the displayed phases at each layer.

1.{\em The charge distribution} is mainly determined by the screening parameter $\alpha$, and is not sensitive to
the temperatures and orders\cite{Okamoto_04, cLin_06}.

2.{\em The staggered $Q_x$ order} at each layer follows the bulk behavior at the corresponding layer charge density density.

3.{\em The uniform $Q_z$ order} is caused by two sources. First (bulk effect), a uniform $-Q_z$ order is induced
with the presence of the staggered $Q_x$ order. Second (SL effect), the charge inhomogeneity induces an
electrostatic like force pushing oxygens closer to the layer with lower density. Note that the origin of this
force is the Mn-O antibonding \cite{cLin_08}.

4.{\em The magnetic order}: The inter-layer coupling is FM except that between layers with densities close to 1.

5.{\em The in-plane conductivity} of the SL is essentially the average of the bulk conductivities, i.e.
for a given charge distribution $\{ n_1, n_2, ..n_N\}$, $\sigma_{xx}^{SL} = \bar{\sigma}_{xx}^{bulk} 
=\sum_{i=1}^N \sigma_{xx}^{bulk} (n_i) /N$ while we do not have simple rule for out-of-plane conductivities.

\subsection{Determining the charge distribution and screening parameter}
\begin{figure}[htbp]
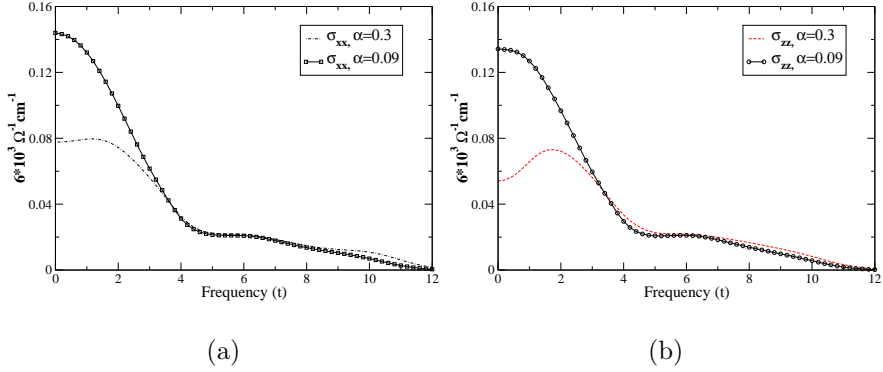

\centering
   \subfigure[]{\epsfig{file = Cond_xx_L2S1_dia12and40_t1.0_t1-0.00_eps0.20_JAF0.020_Cubic0.000_UQ1.4_JH2.0_Us1.2_Temp0.120.eps, 
	width=0.35 \textwidth}} 
   \subfigure[]{\epsfig{file = Cond_zz_L2S1_dia12and40_t1.0_t1-0.00_eps0.20_JAF0.020_Cubic0.000_UQ1.4_JH2.0_Us1.2_Temp0.120.eps,
	width=0.35\textwidth}} 
   \caption{In-plane (a) and out-of-plane (b) optical conductivities 
	for L2S1 for $\alpha=0.09$ (with ligands) and 0.3 (without ligands) at $T=0.12t$. 
	}
   \label{fig:L2S1Cond}
\end{figure}

Our results indicate that the key quantity for the SL is the charge distribution and therefore
determining the screening parameter $\alpha$ is very important. In principle the charge distribution can be
determined by measuring the change in manganese valence in the superlattice \cite{Ohtomo_02}, 
but such measurements have not to our knowledge been performed on manganite superlattices.
Here we propose two measurements to constrain
the values of $\alpha$. The first proposal is to grow superlattices with odd number of LaMnO$_3$ layers
(L3S1 for example) and to measure the total magnetization at low temperature. 
If the charge distribution is sharp, then the ferromagnetic  
interface layers are antiferromagnetically
coupled and the layer has zero net magnetization, whereas if the  
charge distribution is broad, the system will simply be ferromagnetic.
This method suffers from a potential experimental disadvantage: lack  
of perfect oxygen stoichiometry could produce ferromagnetism in the La-rich regions even in the  
presence of a sharp charge distribution \cite{Adamo_08, Huang_97}.

The second one is to measure the optical conductivity. The basic idea is that one should be able to decompose 
the in-plane optical conductivities into sum of bulk results. 
This measurement is free from the interferences
of the impurity scattering or extra oxygens in the superlattice. 
Fig(\ref{fig:L2S1Cond})(a) shows the in-plane L2S1 optical conductivities $\sigma^{L2S1}_{xx}(\omega)$ for $\alpha=0.3$ and $\alpha = 0.09$ at $T=0.12t$ ($\sim$300K and with interfaces being slightly ferromagnetic). We see that for
$\alpha=0.3$, the $\sigma^{L2S1}_{xx}(\omega)$ is not peaked at zero frequency because 1/3 of the contribution
is from Mn layers with density close to one. On the other hand for $\alpha=0.09$, $\sigma^{L2S1}_{xx}(\omega)$ behaves
more like a bulk material. 


The out-of-plane conductivities contain information of the potential differences at different layers.
As discussed in section IV.A, as for the $\sigma_{zz}$ is concerned, the main effect of the layer-dependent 
potentials is to produce layer-dependent local spectral functions which makes $\sigma_{zz}$ different
from $\sigma_{xx}$ at high T non-ordered phase. Therefore the anisotropy in $\sigma$ above ordering temperatures is a very straightforward 
estimator of how confined/delocalized the electrons are. Fig(\ref{fig:L2S1Cond})(b) shows the out-of-plane conductivity 
$\sigma^{L2S1}_{zz}(\omega)$ for the (2,1) superlattice. As expected larger $\alpha$ results in larger anisotropy in $\sigma$. 
For the numerics, because the potential difference is not directly related to the peak in the out-of-plane
optical conductivity as in the model studied in Ref \cite{cLin_06}, one has to compare 
the experimental data with theoretical results.



\subsection{Orbital order proximity effects}
Our calculation indicates that orbital order associated with the $Q_z$ mode (octahedral distortion
with long bond pointing along z direction) is reasonably efficiently  
transmitted from one layer to the next, whereas orbital order associated with $Q_x$ (octahedral  
distortion with long axis in x-y plane) is not transmitted at all. The  
approximations made in this paper amount to retaining only nearest neighbor interactions in both the  
electronic and lattice sectors; in this approximation rotational invariance about the bond in the z direction ensures that  
$Q_x$ order is not transmitted. In this section we estimate the extent to which terms not included in  
our model may change this result.

The transmission of $Q_x$ orbital order implies an energy difference between
states in which $Mn$ ions in adjacent planes have the same or opposite amplitudes
for $Q_x$ order. To estimate this we consider the energy difference between $(\pi,\pi,0)$ and
$(\pi,\pi,\pi)$ orbital order. There are two contributions to this energy: electronic and elastic.

We begin with the electronic term. A given lattice distortion selects one locally favored state
of each site $i$, which we refer to as $\theta_i^g>=cos\theta_i |3z^2- r^2> +sin\theta_i |x^2-y^2>$
and also the orthogonal excited state $\theta_i^e>$.  For the $+Q_x$   
distortion the locally favored state is $|\theta^g_{Q_x}>=\frac{1}{\sqrt{2}}|3z^2-r^2> +\frac{1}{\sqrt{2}}| 
x^2-y^2>$ and the locally disfavored state is $\theta^e_{Q_x}>=\frac{1}{\sqrt{2}}|3z^2-r^2> -\frac{1}{\sqrt{2}}|x^2- 
y^2>$; for the $-Q_x$ distortion the roles of $\theta_g$ and $\theta^e$ are reversed.

The $(\pi,\pi,0)$ $Q_x$ and $(\pi,\pi,\pi)$ $Q_x$ orders produce  
exactly the same local distortions; an electronic contribution
to the energy difference between these two orders can therefore only  
arise from a difference in the electronic hybridization energies for
the two states. We now argue that if only nearest-neighbor hopping is  
present, there can be no hybridization energy difference between
the two states. The in-plane hopping is trivially the same for the  
two states; a difference can therefore only arise from the
$z$ direction hopping.  The hybridization is given by a $2\times 2$  
matrix; in the nearest neighbor hopping approximation the matrix is
degenerate in all three directions with only one non-zero eigenvalue, $t$. For hopping in  
the $z$ direction the only non-vanishing matrix element is
the one connecting $|3z^2-r^2>$ orbitals on the two sites implying  
that the matrix elements between any combination of $|\theta^g_{Q_x}>$
and$ |\theta^e_{Q_x}>$ is $t/2$, so that the nearest neighbor hopping  
does not distinguish between the two states. Including the 2nd NN couplings can lift this degeneracy. 
The energy difference can be estimated by comparing the energy gain caused by the virtual
processes involving 2nd NN hoppings (super-exchange) and is done in Ref \cite{cLin_08}. The LDA calculation \cite{Ederer_07}
suggests the 2nd NN hopping is roughly 0.035eV, leading to an energy difference of the order of meV for these two orders.

We next consider contributions arising from elastic forces.
In a ball and spring model, we have found that oxygen-oxygen or
three-body forces are required to propagate $Q_x$ order. For these
situations, spring constant models are not reliable and band theory
calculations of phonon stiffness are required. Available information on
these forces is presented in \cite{AhnMillis_01}; the result is that the  
energy difference is of order meV, roughly the same magnitudes as
the electronic contribution. With these estimates we conclude that our model captures the
main physics of manganites and the proximity effect for the planar staggered $Q_x$ order
is indeed very weak in reality.

\section{Connections to Experiments}
The temperature dependence of the DC resistivity ($\rho$) and magnetization ($M$) of (LaMnO$_3$)$_{2n}$(SrMnO$_3$)$_{n}$ 
(the (2n,n) superlattice in our notation) have been measured  \cite{Adamo_08, May_07}. In this section 
we discuss the connection between our calculation and these measurements.

\subsection{DC Resistivity}
\begin{figure}[htbp]
   \centering
   \epsfig{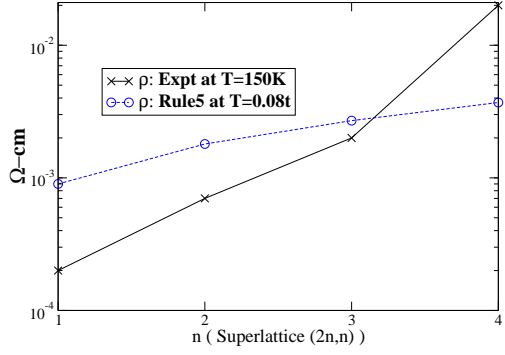}
   \caption{ The in-plane DC resistivities from experiment at $T=150K$ \cite{May_07} (solid line) and
	from Rule 5 at $T=0.08t$ with sharp charge distribution. 
	}
   \label{fig:SL_expt_rule}
\end{figure}

For the resistivity measurement, as the temperature is decreased from 400K, the resistivity $\rho$ is  
found to increase until a superlattice-dependent temperature $T_d$ is reached. This ``downturn'' in $\rho(T)$
is observed for all $(2n,n)$ superlattices. The temperature at which the downturn occurs ($T_d$) is $n$ dependent 
(larger $n$ lower $T_d$) and roughly 200-350$K$ (Fig(2) (a)-(c) in \cite{Adamo_08}, Fig(2) in \cite{May_07}). 
Above $T_d$, the magnitude of $\rho$ has a very strong dependence on $n$.
When the temperature is further reduced below 100$K$, the resistivity increases 
again for $n>3$ superlattice while keeps on decreasing for $n<3$. 

Calculation finds two typical behaviors corresponding to weakly and strongly bound charge distributions
For both cases we expect a downturn in resistivity which is associated with the interface FM/PM transition with
the downturn temperature essentially $n$ independent. At very low temperature our theory predict metallic ($d\rho/dT>0$)
behavior. The $n$ dependence on the magnitude of $\rho$ are different for these two limits:
for the weakly bound charge distribution $\rho$ has very weak $n$ dependence while for the strongly bound
$\rho$ is proportional to the interface density and thus $\propto n$.


The results according to our rules are in reasonable agreement with the $n<3$ superlattices
because they remain metallic in the low temperature. For long period superlattice except the existence of the changing sign in $d\rho/dT$,
our theory is not compatible with experiments at quantitative level because of the exhibited 
$n$-dependence on $\rho$ and $T_d$, and because of the low T insulating phase.
Essentially the experimental measured $\rho$ depends too strongly on $n$ compared to our
theoretical prediction as shown in Fig(\ref{fig:SL_expt_rule}).
We think these inconsistencies are caused by the $n$-dependent interface quality.



\subsection{Magnetization}

For the magnetization measurement, Ref\cite{Adamo_08} (Fig(2) (d)-(f)) shows that 
all superlattices exhibit a net magnetization below some onset temperature $T^I_c$ (the superscript $I$ denotes interface).
$T^I_c$ depends on $n$ (decrease in $T^I_c$ from 300K to 200 when increasing $n$ from 2 to 16 ) 
and is very close to the downturn temperature in resistivity $T_d$.
For the $n=2$ superlattice, the magnetization increases and saturates when lowering the temperature.
For $n=4$ $n=16$ superlattices, the magnetization increases slowly right below $T^I_c$, then more rapidly
around 150K. When the system is further cooled without applying a magnetic field (zero field cooling, ZFC), 
the magnetization reaches a maximum (roughly the same value $\sim 0.4 \mu_B$/Mn for both $n=4$ and $n=16$) 
around 110K, before eventually decays to zero. 
If the system is cooled in an applied magnetic field (field cooling, FC), the $n=2$ superlattice behavior is not changed
but for both n = 4 and n = 16 the magnetization saturates at 0.6$\mu_B$/Mn approaching zero temperature.
We believe that the FC measurements reveal more clearly the intrinsic behavior and will therefore
focus on the FC data but will comment on the ZFC data in the end.


Our calculation predicts that ferromagnetism is associated with the interfaces.
There is a critical thickness which depends on the width of the charge distribution (value of $\alpha$).
For systems with $n$ greater than this critical value (which may be 0   
for very sharp charge distributions), the magnetization resides only on the interfaces,
so one expects an onset at the interface $T_c$ (comparable to the  
bulk value) and a saturation value
scaling with the interface density (with coefficient depending on $\alpha$).

Now we discuss the magnetization (the FC data) based on our calculation. First our theory agrees reasonably well
with the $n=2$ (L4S2) superlattice assuming the critical thickness is $\gtrsim 3$ so the LaMnO$_3$ layers
are ferromagnetically ordered (the  $n=2$ superlattice saturates at 2.8$\mu_B$/Mn at low temperature indicating 
LaMnO$_3$ layers are ferromagnetic).
However, the behavior of the $n=4$ and $n=16$ superlattices is not 
consistent with our theory, except in the qualitative sense
that that resistivity downturn and magnetic onset temperatures coincide.  The strong $n$-dependence of the
magnetism-onset temperature $T_c^I$ is not found in the calculation, where ferromagnetism is an interface phenomenon and therefore
not strongly $n$-dependent. Further, the observed rapid growth in $M$  
below $140K$ and the saturation at about $1$ $\mu_B$ displayed in the  
long period ($n>6$) superlattices is not found in our theory, which instead  
implies {\it antiferromagnetism} below $140K$ and a saturation
magnetization proportional to the interface density.


We believe that the inconsistency is a consequence of oxygen stoichiometry that extra oxygens reside
in La-riched regions so that LaMnO$_3$ is effectively a slightly doped manganites \cite{Huang_97,Adamo_08}.
Based on this assumption and our rules, we now explain the FC data. Our rule states that each layer behaves
as a bulk manganite at the given charge density. For the density close to but not equal to one, 
the system displays canted A-AF order \cite{CMR} around 140K which has a small but non-zero net magnetic moment. 
Within this picture, the constant but small saturation magnetization for $n>6$ superlattices is
the net magnetization from the canted A-AF order, and the rapid growth around 140K which is caused by 
the forming of the canted A-AF order in the La-riched regions. The remaining issue is the $n$ dependence on the interface
Curie temperature. We believe this is caused by the $n$-dependent interface quality. Since the main energy gain for 
PM/FM transition is the kinetic energy (double-exchange), we expect the bad interface quality tending to localize electrons 
will reduce the Curie temperature. This is qualitatively consistent with the observation.

Finally we comment on the zero field cooled data. The most striking feature for ZFC data is that for the $n\geq 4$ superlattice
the magnetization starts to decrease around 110K upon cooling. We associate this reduction in $M$ with the G-AF order developed
in the SrMnO$_3$ layers. The forming of G-AF order in SrMnO$_3$ weakens or temporarily kills the ferromagnetic coupling
between ferromagnetic interfaces therefore reduces the total magnetization. Note that the Neel temperature for
CaMnO$_3$ is 110K \cite{Wollan_55}.

\subsection{Interface Qualities and Proposed Experiments}
Although the qualitative interpretations can be made,
our rules are not compatible with experimental data at quantitative level, in particular the resistivity measurement.
We  believe this discrepancy is mainly caused by the $n$-dependent interface quality which has two possible causes. 
The first one is the mismatch in lattice constant. Since the LaMnO$_3$ and SrMnO$_3$ have slightly different
lattice constants, the interface either tolerates very large strain forces or more probably creates some defects to
compensate the mismatch. The presence of defects increases the scattering sources and thus the resistivity which is not
considered in our model. The second one is the magnetic frustration. Since LaMnO$_3$, SrMnO$_3$ at low temperature
display A-type and G-type AF orders respectively, interfaces undergo the magnetic frustration which
suppresses the tendency of FM order and again increases the resistivity.

To reduce these extra variables in experiments, we propose one should conduct experiments on $(n,1)$ SL.
For this $(n,1)$ series SL, since no Mn are sandwiched by Sr layers we expect
the effects due to the lattice constant mismatch should be relatively small. Perhaps more importantly 
complications from the magnetic frustration caused by SrMnO$_3$ AFM regions will not enter the problem.
Based on these arguments the $n$ dependent interface quality is expected to be strongly reduced and the
direct comparison between experiments and theory becomes easier.

\section{Conclusion}
We have studied the (LaMnO$_3$)$_n$(SrMnO$_3$)$_1$ $(n,1)$ superlattice. We found that the charge distribution,
determined by the screening parameter $\alpha$, is the key quantity and propose that the optical conductivity measurement of 
L2S1 can fix it. Once the charge density at each layer
is known, the phases essentially follows the bulk at the corresponding density. General rules for phases
are given. We propose the measurements on low temperature magnetization of the L3S1 SL and  on the optical conductivity 
help constrain the range of $\alpha$. 
For $\alpha=0.3$ which results in sharp charge distribution, we predict at low temperature the $(n,1)$ superlattice
has no net magnetization with period $2\times(n+1)$ for odd $n$. Finally we comment on recent experiments based 
on our calculations and infer that the main discrepancy is caused by the sample-dependent interface quality.
We suggest the $(n,1)$ superlattice is actually a physically simpler and cleaner system to study.

\section{Acknowledgment}
We acknowledge support from DOE-ER46189 and the Columbia MRSEC.




\end{document}